# Modeling the Internet's Large-Scale Topology

Soon-Hyung Yook, Hawoong Jeong, Albert-László Barabási

*Department of Physics, University of Notre Dame, Notre Dame, IN 46556, USA*

**Network generators that capture the Internet's large-scale topology are crucial for the development of efficient routing protocols and modeling Internet traffic[1,2,3]. Our ability to design realistic generators is limited by the incomplete understanding of the fundamental driving forces that affect the Internet's evolution. By combining several independent databases capturing the time evolution, topology and physical layout of the Internet, we identify the universal mechanisms that shape the Internet's router and autonomous system level topology. We find that the physical layout of nodes form a fractal set[4,5], determined by population density patterns around the globe. The placement of links is driven by competition between preferential attachment and linear distance dependence, a marked departure from the currently employed exponential laws[6,7]. The universal parameters that we extract significantly restrict the class of potentially correct Internet models and indicate that the networks created by all available topology generators are fundamentally different from the current Internet.**

In the light of extensive evidence that Internet protocol performance is greatly influenced by the network topology[1,2,3], network generators are a crucial prerequisite for understanding and modeling the Internet. Indeed, security and communication protocols perform poorly on topologies offered by generators different from which they are optimized for, and are often ineffective when released[2]. Protocols that work seamlessly on prototypes fail to scale up, being inefficient on the larger real network[1]. Thus in order to efficiently control and route traffic on an exponentially expanding Internet[8], it is important that topology generators not only capture the structure of the current Internet, but allow for efficient planning and long term network design as well[1].

Our ability to design good topology generators is limited by our poor understanding of the basic mechanisms that shape the Internet's large scale-topology. Indeed, until recently all Internet topology generators[6,9,10] offered versions of random graphs[11,12]. The 1999 discovery of Faloutsos *et al.*[13] that the Internet is a scale-free network with a power-law degree distribution[14] invalidated all previous modeling efforts. Subsequent research confirmed that the difference between scale-free and random networks are too significant to be ignored: protocols designed for random networks fare poorly on a scale-free topology[2]; a scale-free Internet displays high tolerance to random node failures but is fragile against attacks[15,16,17]; computer viruses spread threshold free on scale-free networks[18] with obvious consequences on network security. These insights motivated the development of a new brand of Internet topology generators[7,19] that offer scale-free topologies in better agreement with empirical data. Despite these rapid advances, it is unclear that we are aware of all driving forces that govern the Internet's topological evolution. It is therefore of crucial importance to perform measurements that directly probe and uncover the mechanisms that shape the Internet's large-scale topology.

Here we offer direct experimental evidence for a series of fundamental mechanisms that drive the Internet's evolution and large-scale structure. In contrast with the random placement of nodes, we find that the Internet develops on a fractal support, driven by the fractal nature of population patterns around the world. In contrast with current modeling paradigms, that assume that the likelihood of placing a link decays exponentially with the link's length, we find that this dependence is only linear. Finally, we offer quantitative evidence that preferential attachment, responsible for the scale-free topology, follows a linear functional form on the Internet. These results allow us to identify a class of models that could serve as a starting point for topologically correct network generators. Surprisingly, the obtained phase diagram indicates that all current Internet network generators are in a different region of the phase space than the Internet.

At the lowest resolution the Internet is a network of routers connected by links. As each router belongs to some administrative authority, or autonomous system (AS), the Internet is often considered as a network of interconnected ASs. For completeness, here we study simultaneously the router and AS level topology, using the term *node* to represent both routers and ASs, unless specified otherwise.

*Physical layout:* Current Internet topology generators assume that routers and domains are distributed *randomly* in a two dimensional plane[6,7,9,10,19]. In contrast, we find that routers and ASs form a fractal set[4], strongly correlating with the population density around the world. In Fig. 1a we show a map of the worldwide router density, obtained using the NetGeo tool to identify the geographical coordinates of 228,265 routers provided by the currently most extensive router-level Internet mapping effort[20]. Compared with the population density map (Fig. 1b), the results indicate strong, visually evident correlations between the router and the population density in economically developed areas of the world. We used a box counting method[4,5] to analyze the spatial distribution of router, domain, and population density. The results, shown in Fig. 2a, indicate that each of the three sets form a fractal with dimension $D_f$ =1.5 ± 0.1. The coincidence between the fractal dimension of the population and the Internet (router and AS) nodes is not unexpected: high population density implies higher demand for Internet services, resulting in higher router and domain density. Fig. 2b supports the existence of such correlations, indicating that the router and AS density increase monotonically with the population density.

*Placing the links:* Connecting two nodes on the Internet requires extensive resource and time investment. Thus network designers prefer to connect to the closest node with sufficient bandwidth, a process that clearly favors shorter links. To discourage long links, all topology generators are based on the Waxman model[6], which assumes that the likelihood of placing a link between two nodes separated by the Euclidean distance *d* decays as *P(d) ~ exp(-d/d₀)*, where $d_0$ is a free parameter taken to be proportional to the system size. Despite its wide use in Internet topology generators[6,7,9], there is no empirical evidence for such exponential form, which forbids links between far away nodes. Intuition suggests otherwise: one would expect that the likelihood of connecting two nodes is inversely proportional with the distance between the nodes, i.e. *P(d)~1/d*. The correct functional form of *P(d)* is crucial for Internet modeling: our simulations indicate that a network developing under the Waxman rule asymptotically converges to a network with exponentially decaying degree distribution, in contrast with the power-law documented for the Internet. Therefore, to uncover the proper form of *P(d)* we measured the length distribution of the documented Internet links. The results, shown in Fig. 2c, indicate that both router and AS level *P(d)* decays linearly with *d*, siding with our intuition and excluding Waxman's rule.

*Preferential attachment:* Preferential attachment is believed to be responsible for the emergence of the scale-free topology in complex networks[14]. It assumes that the probability that a new node will link to an existing node with *k* links depends linearly on *k*, i.e. $\Pi(k) = k / \sum_i k_i$. On the other hand, in real systems preferential attachment could have an arbitrary nonlinear form. Calculations indicate, however, that for $\Pi(k) \sim k^\alpha$, with $\alpha \neq 1$ the degree distribution deviates from a power law[21]. In the light of these results, in order to properly model the Internet, we need to determine the precise functional form of $\Pi(k)$. To achieve this, we use Internet AS maps recorded at 6 months intervals, allowing to calculate the change $\Delta k$ in the degree of a AS node with *k* links during the investigated time frame. The results indicate that the rate at which a node increases its degree is linearly proportional with the number of links the node has, offering quantitative support for the presence of *linear* preferential attachment (Fig. 2d), supported by independent measurements as well[21].

Taken together, our measurements indicate that four mechanisms, acting independently, contribute to the Internet's large-scale topology. First, in contrast with classical network models the Internet grows incrementally, being described by an evolving network[14,21,22,23] rather than a static graph[11,12]. Second, nodes are not distributed randomly, but both routers and domains form a scale-invariant fractal set with fractal dimension $D_f = 1.5$. Finally, link placement is determined by two competing mechanisms. First, the likelihood of connecting two nodes decreases linearly with the distance between them, and second, the likelihood of connecting to a node with *k* links increases linearly with *k*. Building on these mechanisms, each supported independently by our measurements, we propose a general model that offers an integrated framework to investigate the effect of the different mechanisms on the Internet's large-scale topology.

Consider a 'map', mimicking a continent, which is a two dimensional surface of linear size $L$. The map is divided into squares of size $\ell \times \ell$ ($\ell \ll L$), each square being assigned a population density $\rho(x,y)$ with fractal dimension $D_f$. At each time step we place a new node $i$ on the map, its position being determined probabilistically, such that the likelihood of placing a node at $(x,y)$ is linearly proportional with $\rho(x,y)$. We assume that the new node connects with $m$ links to nodes that are already present in the system. The probability that the new node links to a node $j$ with $k_j$ links at distance $d_{ij}$ from node $i$ is

$$\Pi(k_j, d_{ij}) \sim k_j^{\alpha}/d_{ij}^{\sigma}, \qquad (1)$$

where $\alpha$ and $\sigma$ are pre-assigned exponents, governing preferential attachment and the cost of the node-node distance. Increasing $\alpha$ will favor linking to nodes with higher degree, while a higher $\sigma$ will discourage long links.

The parameters of the model can be assigned into two qualitatively different classes. First, $L$, $\ell$ and $m$ are non-universal parameters, as their value can be changed without affecting the network's large-scale topology. On the other hand, $\alpha$, $\sigma$, and $D_f$ are universal exponents, as their values uniquely parameterize a family of Internet models, generating potentially different large-scale topologies. Therefore, we use a three dimensional phase-space whose axis are the scaling exponents, $2-D_f$, $\alpha$ and $1/\sigma$ (Fig. 3) to identify the possible scaling behavior predicted by the model. For easy reference, we show the location within this phase space of all currently used Internet topology generators. Our measurements (Fig. 2) allow us to unambiguously identify the position of the Internet within this phase space at $\sigma=1$, $\alpha=1$ and $D_f=1.5$, clearly separated from all network generators. Such separation should not be a problem if some of the models and the Internet belong to a region of the phase space that share the same universal topological features. We will show next, however, that this is not the case, as any deviation from the point denoting the Internet significantly alters the network's large-scale topology.

To systematically investigate the changes in the network topology as we deviate from the point denoting the Internet next we consider the effect of changing $\sigma$, $\alpha$, and $D_f$, moving separately along the three main axis.

*Varying $\sigma$* while leaving $\alpha=1$ and $D_f=1.5$ unchanged changes the contribution of the Euclidean distance to the network topology, interpolating between the $\sigma=0$ phase corresponding to the scale-free model and the $\sigma=\infty$ limit, corresponding to the Waxman rule. As Fig. 4a shows, for an exponential distance dependence (Waxman's rule) the degree distribution $P(k)$ develops an exponential tail, disagreeing with the power law $P(k)$ of the Internet[13,20]. Moving towards the $\sigma=0$ axis, as the physical distance gradually looses relevance in (1), we recover the scale-free model, for which the physical layout does not influence the network topology. Changing $\sigma$ affects the link length distribution $P(d)$ as well. As shown in Fig. 4b, for $\sigma=1$ the $P(d)$ distribution quantitatively agrees with the *1/d* dependence uncovered by the direct measurements (Fig. 2c), but in the exponential limit of the Waxman rule we find that $P(d)$ develops an exponential tail, a functional form that characterizes the full $1/\sigma=0$ plane (i.e. valid for arbitrary $\alpha$ and $D_f$ as long as $\sigma=\infty$). Finally, decreasing $\sigma$ does not seem to affect $P(k)$, but it does change $P(d)$, which for $\sigma=0$ develops an extended plateau for $d/R<10^{-2}$, followed by rapid decay, in disagreement with the measurements (Fig. 2c). Consequently, moving with $\sigma$ away from the empirically determined $\sigma=1$ value affects both the degree and the distance distribution, creating significant deviations from the known Internet topology.

*Varying $\alpha$* while leaving $D_f$ and $\sigma$ unchanged has drastic immediate effects on the degree distribution, forcing on it an exponential form. Indeed, for any $\alpha<1$, the distribution $P(k)$ develops an exponential tail, turning into a pure exponential for $\alpha=0$, when preferential attachment is absent (Fig. 4c). This agrees with the analytical predictions of Kapriwsky and Redner[21], who have shown that $\alpha<1$ destroys the power law nature of the degree distribution in the scale-free model. The calculations predict that for $\alpha>1$ gelation takes place, leading to a network architecture in which all nodes are connected to a central node. Our simulations fully confirm these prediction in the vicinity of the $D_f=1.5$ and $\sigma=1$ point, as for $\alpha>1$ the $P(k)$ distribution develops an elongated non-power law tail, corresponding to a few highly connected nodes, a characteristic feature of gelation. Furthermore, our simulations indicate that the gelation phase is present in the full $\alpha>1$ region of the phase space, colored yellow in Fig. 3. Consequently, we find that any deviation from $\alpha=1$ results in a significant alternation of the network's *P(k)* distribution, while having little effect on *P(d)* (Fig. 4d).

*Varying $D_f$* while leaving $\sigma$ and $\alpha$ unchanged has little visible effect on the degree distribution (Fig 4e). The only changes appear in *P(d)*. Indeed, we find that placing the nodes proportional to the population density, thereby creating a fractal with $D_f = 1.5$, results in a *P(d)* distribution in agreement with the data points provided by the direct measurements (Fig 4f). On the other hand, a random node distribution, corresponding to $D_f = 2$, generates a *P(d)* that is visibly different from the real data. A uniform distribution appears to lead to a long plateau in *P(d)*, followed by a faster decay than seen in the real Internet.

In summary, our measurement and simulations indicate that the Internet takes up a very special point in the ($\alpha$, $D_f$, $\sigma$) phase space, such that any deviation from its current position identified by direct empirical measurements significantly alters the network topology. Interestingly, we find that all current generators lay in a different region of the phase space than the Internet (Fig. 3), indicating that they generate networks that belong to a different topological class. While the changes induced by not considering the fractal nature of the router distribution are less striking, we find that the use of $\sigma \neq 1$, a feature of all available network generators, has drastic topological consequences.

Identifying the location of the Internet within this phase space does not automatically offer an Internet model valid to ultimate details, as there are several non-universal characteristics that contribute to the network topology. For example, several studies have established that the value of the degree exponent $\gamma$ can be tuned by changing the model parameters[14,21,22,23], as the relative frequency of node and link addition and removal jointly determine $\gamma$. To predict the value of the degree exponent one needs to carefully measure all frequencies and include internal links, which requires time resolved Internet maps that are currently available only at the low resolution AS level only. Consequently, the degree exponent can take up any value, while leaving the Internet's position in the phase space unchanged. Similarly, the Internet displays nontrivial higher order correlations, that could be explained by incorporating node fitness[24]. Therefore, to design topology generators that reproduce simultaneously the precise numerical values and the correct functional form of the Internet's path length and degree distribution, one needs to incorporate numerous Internet specific details. However, our results indicate that several universal constraints influence the network's large-scale topology. That is, no matter how detailed an Internet model is, if its universal parameters ($\alpha$, $\sigma$, $D_f$) deviate from those uncovered by measurements, the large-scale topology will inevitably differ from the current Internet.

The advantage of the model proposed here is its flexibility: it offers an universally acceptable skeleton for potential Internet models, on which one can build features that could lead to further improvements. Using an evolving network to model the Internet has the potential to predict the future of the network, as the model incorporates only time invariant mechanisms that should continue driving the network's development in the future. Such predictive power, combined with elements pertaining to link bandwidths and traffic predictions[25] could offer a crucial tool to uncover potential bottlenecks and network congestions resulting from the Internet's rapid, decentralized development. These advances are crucial for both scientific and design purposes, being a key prerequisite for developing the next generation communication technologies.

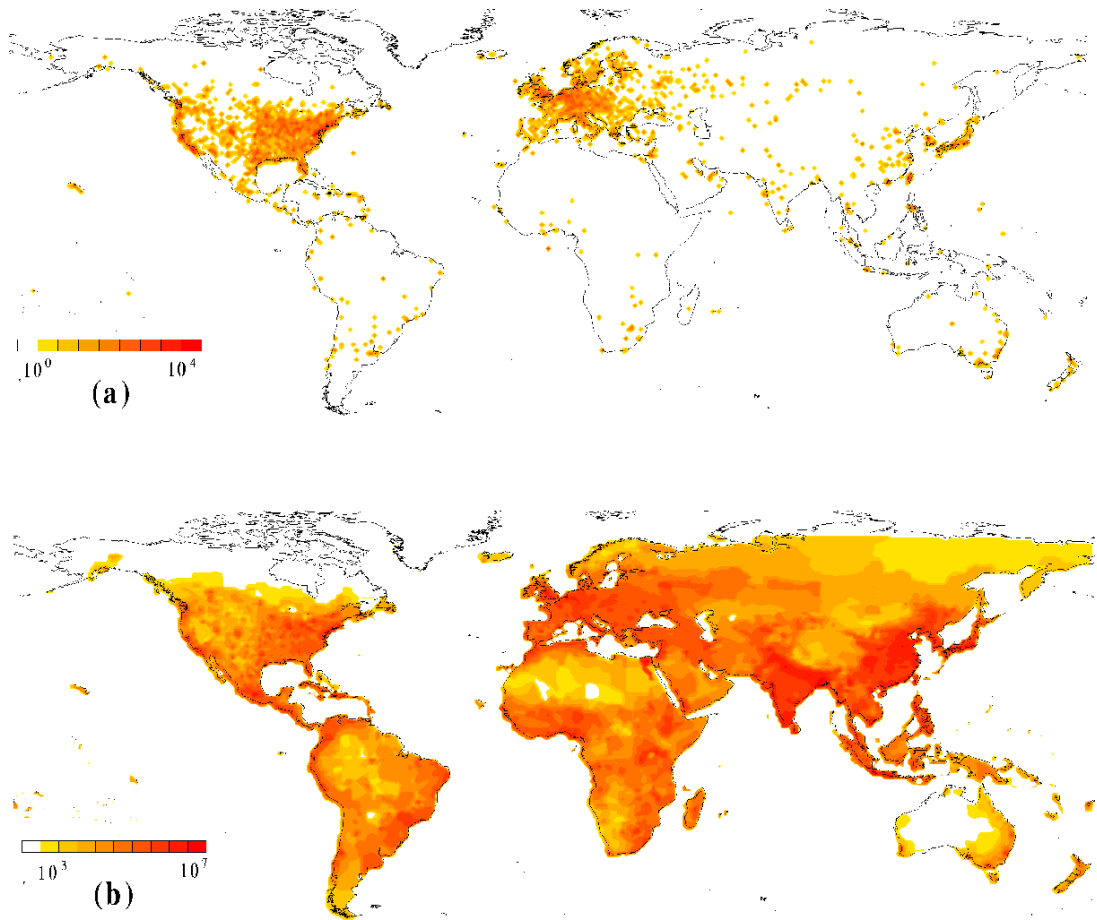

**Figure 1. Distribution of the Internet around the world.** (**a**) Worldwide router density map obtained using NetGeo tool (http://www.caida.org/tools/utilities/netgeo) to identify the geographical location of 228,265 routers mapped out by the extensive router level mapping effort of Govindan and Tangmunarunkit[20]. (**b**) Population density map based on the CIESIN's population data (http://sedac.ciesin.org/plue/gpw). Both maps are shown using a box resolution of 1°×1°. The bar next to each map gives the range of values encoded by the color code, indicating that the highest population density within this resolution is of the order $10^7$ people/box, while the highest router density is of the order of $10^4$ routers/box. Note that while in economically developed nations there are visibly strong correlations between population and router density, in the rest of the world Internet access is sparse, limited to urban areas characterized by population density peaks.

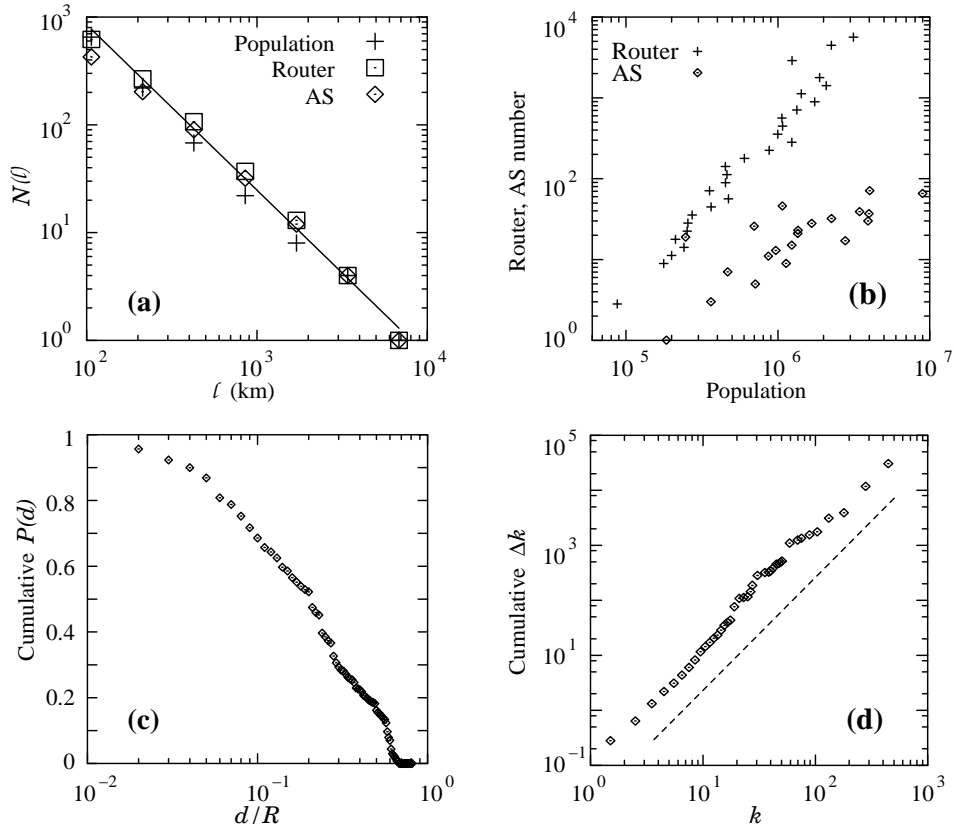

**Figure 2. Characterizing the Internet's physical layout and topology using direct measurements.** (**a**) The physical layout of the Internet was studied using box counting method[4,5], applied to the map shown in Figure 1. The log-log plot shows the number of boxes of size $\ell \times \ell$ km with nonzero routers/AS/inhabitants in function of $\ell$ for North America. The slope of the straight line indicates that $D_f \approx 1.5 \pm 0.1$ for each dataset. (**b**) The dependence of the router/AS density on the population density in North America, showing the average number of router/AS nodes in a $1° \times 1°$ box in function of the number of people living in the same area. Similar plots were obtained for each continent, the steepness of the curves strongly correlating with economic factors. To determine the AS density we used 12,409 ASs from NAI (http://moat.nlanr.net /infrastructure.html), combined with NetGeo to identify their geographical location. (**c**) The length distribution of the links connecting routers, shown as a function of the dimensionless variable $d/R$, where $d$ is the Euclidean distance between two routers and $R$=6,378 km is the radius of the Earth. The linearly decaying cumulative $P(d)$ on a log-linear plot indicates that $P(d) \sim 1/d$. We removed the first point, corresponding to $d/R \approx 0$, from the figure, as that collects within a single box all router pairs that our map resolution does not resolve, creating an artificially large router density. Higher resolution maps should automatically eliminate this artifact. (**d**) The cumulative change $\Delta k$ in the connectivity of AS nodes with $k$ links. The dotted line has slope 2, indicating that the cumulative $\Delta k \sim k^2$, i.e. the change in $\Delta k$ is linear in $k$, offering direct proof for linear ($\alpha$=1) preferential attachment.

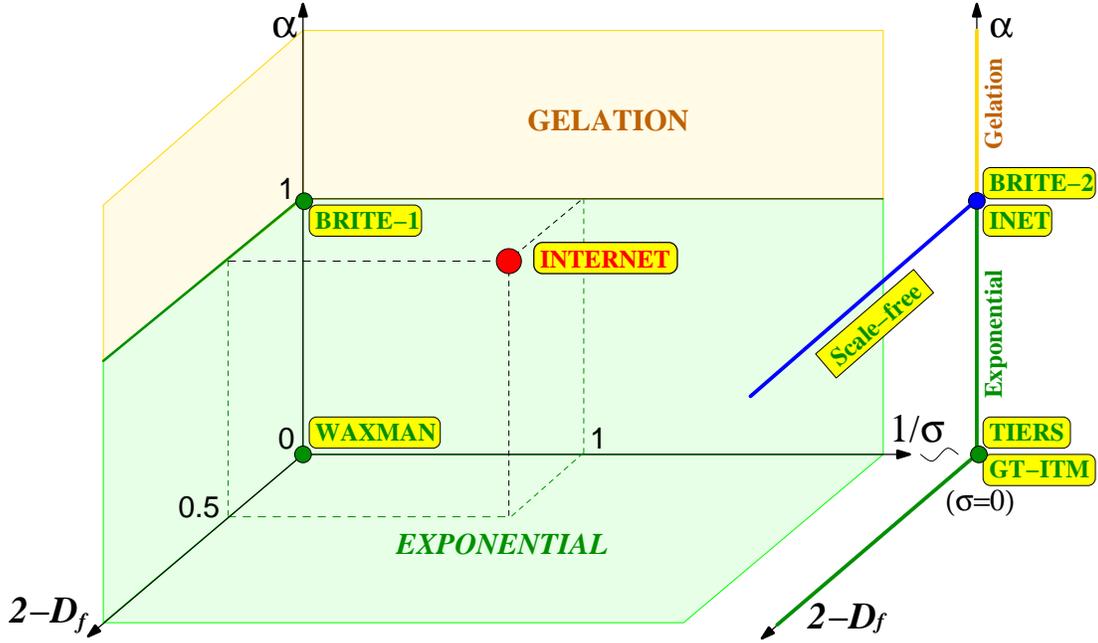

**Figure 3. Phase diagram summarizing various Internet models and their expected large-scale topology.** The axes represent the scaling exponents, $1/\sigma$, $2-D_f$ and $\alpha$, governing, respectively, link placement, node location and preferential attachment. As several models of interest are at $\sigma=0$, which is at $1/\sigma=\infty$, at the right we added a separate axis marking this special point in the phase space. Our measurement indicate (Fig.2) that within this phase space the Internet can be found at $1/\sigma=1$, $2-D_f=0.5$, and $\alpha=1$, identified as a red circle. The yellow boxes indicate the location of all current Internet topology generators. WAXMAN[6]: Nodes are placed randomly in space ($D_f=2$) with exponential distance dependence ($1/\sigma=0$) and without preferential attachment ($\alpha=0$). TIERS[8]: Based on a three level hierarchy the model has no space dependence ($D_f=2$, $1/\sigma=\infty$) and no preferential attachment ($\alpha=0$). GT-ITM[9]: While based on several different models, the most used PureRandom transit-stub version occupies the same position in the phase space as TIERS ($D_f=2$, $1/\sigma=\infty$, $\alpha=0$). INET2.0[19] connects randomly placed nodes using an externally imposed power-law connectivity information. While it does not include preferential attachment explicitly, as $P(k)$ is forced to follow a power-law, we put this generator at ($D_f=2$, $1/\sigma=\infty$, $\alpha=1$). Note, however, that there are significant known differences[26] between a static graph, such as generated by INET, and scale-free topologies generated by evolving networks, such as the Internet. BRITE[7]: The most advanced of all, BRITE incorporates preferential attachment ($\alpha=1$) combined with the Waxman rule ($1/\sigma=\infty$) for placing the links. As BRITE has the option to produce topologies with different parameters, we denote by BRITE-1 the version with only preferential attachment ($D_f=2$, $1/\sigma=\infty$, $\alpha=1$), and BRITE-2 the version including the Waxman rule as well ($D_f=2$, $1/\sigma=0$, $\alpha=1$). Note that BRITE has as option to include inhomogeneous node placement, creating regions with high node density mimicking highly populated areas. The algorithm, however, does not create a fractal, thus we choose $D_f=2$ for both BRITE-1 and BRITE-2. The scale-free model[14], which ignores the physical location of the nodes ($\sigma=0$ thus $D_f$ can be arbitrary) is shown as a separate blue line on the $1/\sigma=\infty$ axis and $\alpha=1$. The green areas correspond to an exponential $P(k)$ distribution, while yellow areas are characterized by gelation, indicating that the Internet strikes a delicate balance at the boundary of these two topologically distinct phases.

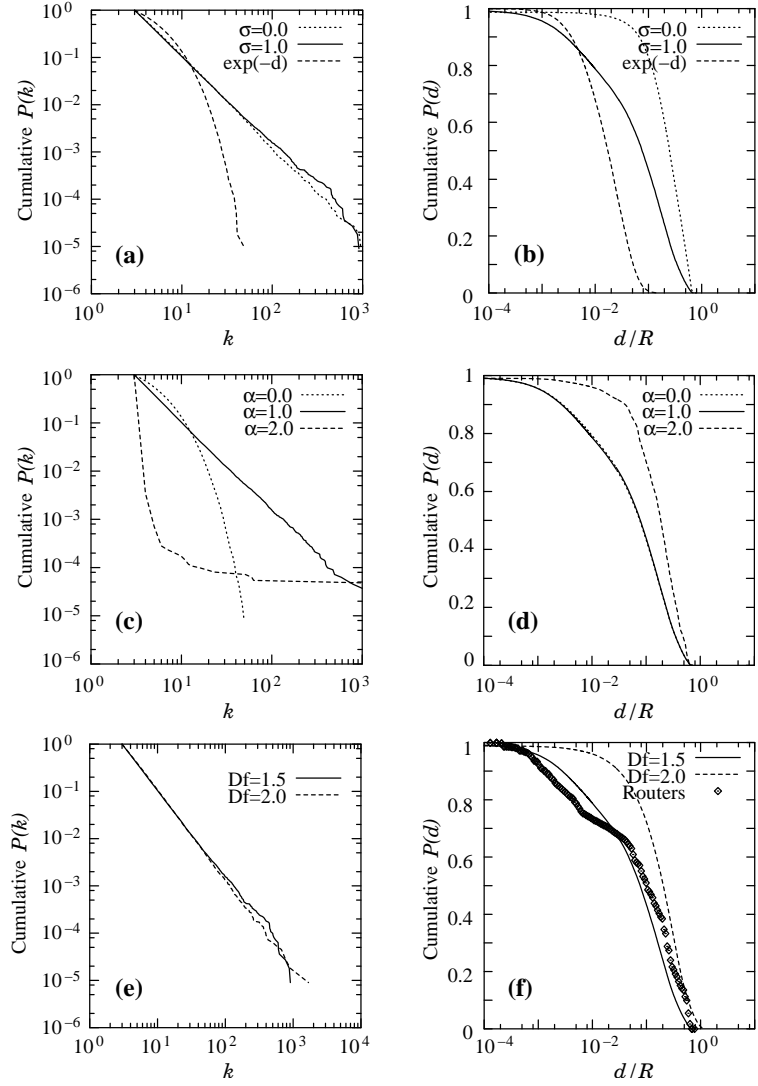

**Figure 4. The dependence of the degree distribution (left column) and link length distribution (right column) on the scaling exponent $\sigma$, $\alpha$, and $D_f$.** (**a**) and (**b**): The effect of changing $\sigma$, with $\alpha=1$ and $D_f=1.5$ unaltered, on $P(k)$ (**a**) and $P(d)$ (**b**). Note that the exponential Waxman rule corresponds to $\sigma=\infty$, in which case $P(k)$ develops an exponential tail. (**c**) and (**d**): The effect of changing $\alpha$ while fixing $\sigma=1$ and $D_f=1.5$ on $P(k)$ (**c**) and $P(d)$ (**d**). For $\alpha<1$ $P(k)$ develops an exponential tail, while for $\alpha>1$ gelation takes place. (**e**) and (**f**): The effect of changing the fractal dimension $D_f$ while fixing $\alpha=1$ and $\sigma=1$ on $P(k)$ (**e**) and $P(d)$ (**f**). Note that changing $D_f$ from a fractal ($D_f=1.5$) to a homogeneous non-fractal distribution ($D_f=2$) leaves $P(k)$ practically unchanged. On the other hand, for $D_f=2$ corresponding to random node placement the $P(d)$ distribution deviates from the data points measured for the Internet, shown as symbols in (**f**). All simulations were carried until $N=109,533$ nodes have been added to the network, which is the size of the currently available router level network maps for North America. We used $m=3$ to match the Internet's known average degree.